\documentclass[conference]{IEEEtran}
\IEEEoverridecommandlockouts
\usepackage{amsmath,amssymb,amsfonts}
\usepackage{algorithm}
\usepackage{graphicx}
\usepackage{textcomp}
\usepackage{xcolor}
\usepackage[noend]{algpseudocode}
\usepackage{booktabs}
\usepackage{url}

\def\BibTeX{{\rm B\kern-.05em{\sc i\kern-.025em b}\kern-.08em
    T\kern-.1667em\lower.7ex\hbox{E}\kern-.125emX}}
\begin{document}

% Revision macro: wraps new/modified text in blue for review.
% To finalize, replace with: 
\newcommand{\rev}[1]{#1}

\title{NetCAS:\\ Dynamic Cache and Backend Device Management in Networked Environments}

\author{
\IEEEauthorblockN{Joon Yong Hwang\thanks{* These authors contributed equally to this work.}\textsuperscript{*}}
\IEEEauthorblockA{\textit{Sungkyunkwan University} \\
\textit{Department of Computer Science} \\
\textit{and Engineering}\\
Suwon, South Korea \\
brian0316@skku.edu}
\and
\IEEEauthorblockN{Chanseo Park\textsuperscript{*}}
\IEEEauthorblockA{\textit{Sungkyunkwan University} \\
\textit{Department of Electrical and} \\
\textit{Computer Engineering}\\
Suwon, South Korea \\
cstim@g.skku.edu}
\and
\IEEEauthorblockN{Younghoon Kim\thanks{\textdagger\ Corresponding author.}\textsuperscript{\textdagger}}
\IEEEauthorblockA{\textit{Ajou University} \\
\textit{Department of Software} \\
\textit{\hphantom{Department of Software}}\\
Suwon, South Korea \\
yhoon@ajou.ac.kr}
}

% \author{
% \IEEEauthorblockN{Anonymous Author\thanks{* These authors contributed equally to this work.}\textsuperscript{*}}
% \and
% \IEEEauthorblockN{Anonymous Author\textsuperscript{*}}
% \and
% \IEEEauthorblockN{Anonymous Author\thanks{\textdagger\ Corresponding author.}\textsuperscript{\textdagger}}

\maketitle

\begin{abstract}
Modern storage systems often combine fast cache with slower backend devices to accelerate I/O. As performance gaps narrow, concurrently accessing both devices, rather than relying solely on cache hits, can improve throughput. However, in data centers, remote backend storage accessed over networks suffers from unpredictable contention, complicating this split. We present NetCAS, a framework that dynamically splits I/O between cache and backend devices based on real-time network feedback and a precomputed Perf Profile. Unlike traditional hit-rate-based policies, NetCAS adapts split ratios to workload configuration and networking performance. NetCAS employs a low-overhead batched round-robin scheduler to enforce splits, avoiding per-request costs. It achieves up to 174\% higher performance than traditional caching in remote storage environments and outperforms converging schemes like Orthus by up to 3.5× under fluctuating network conditions.
\end{abstract}

%TODO: Change later

%-------------------------------------------------------------------------------
\section{Introduction}
%-------------------------------------------------------------------------------

Recent advances in storage technology have narrowed the performance gap between cache and backend storage. Whereas older hierarchies~(e.g., HDD + SSD) showed clear asymmetry, modern pairings~(e.g., NVMe + PMem) often exhibit overlapping throughput and latency. This trend has led to a range of systems that depart from cache-centric hierarchies and instead exploit parallel utilization of heterogeneous devices to maximize throughput. A growing body of work has shown that distributing I/O across tiers, rather than funneling all traffic through the cache, can harness aggregate bandwidth, avoid device-specific bottlenecks, and sustain performance at scale. Recent work in both file systems~\cite{kwon17,zheng19,zhan25,ren25} and caching frameworks~\cite{wu21,lin23,chen21,writes-hurt} demonstrates that parallel, nonexclusive access to multiple devices consistently outperforms strict hierarchical layering.

While prior efforts demonstrated the benefits of parallel utilization, they largely assumed local environments where cache and backend devices are co-located within the same host. Modern datacenter architectures, however, increasingly rely on disaggregated storage~\cite{aws-aurora, azure-hyperscale} designs where application servers access remote storage devices over networks. This separation introduces a key performance challenge: backend device performance becomes unstable and fluctuates due to network variability. Even with RDMA and advanced fabrics, remote I/O remains subject to congestion and interference that can inflate latency and reduce throughput unpredictably~\cite{jin17netcache, shu2023disaggregated, guz2018performance}. As a result, static or converge-based I/O splitting strategies, which implicitly assume stable device performance, are insufficient in these environments, and adaptive network-aware approaches are needed to dynamically adjust caching and splitting decisions in response to observed network and device conditions.

This paper presents NetCAS, a network-aware caching and splitting framework that extends non-hierarchical caching into remote-storage environments. NetCAS addresses the limitations of the prior splitting approaches by introducing a precomputed Perf Profile based dynamic split model. For each workload configuration, NetCAS consults pre-profiled data to determine an ideal split ratio. At runtime, it monitors network performance in real time and adjusts the estimated backend device performance accordingly. This allows NetCAS to dynamically compute a new split ratio that reflects current network conditions.

Through extensive evaluation, we show that NetCAS:
\begin{itemize}
  \item Achieves up to 174\% performance improvement compared to traditional caching in remote storage environments.
  \item Outperforms hit-rate–based converging approaches like OrthusCAS by up to 3.5$\times$ under fluctuating network conditions.
  \item Introduces negligible CPU overhead by integrating directly into OpenCAS’s fast I/O path and avoiding per-request decision-making outside existing control flow.
\end{itemize}
By bridging the gap between adaptive caching and dynamic network conditions, NetCAS offers a scalable and efficient solution for next-generation hybrid storage systems in modern datacenters.

%--------------------------------------------------------------------------
\section{Background and Motivation}
\label{sec:background}
%--------------------------------------------------------------------------
\subsection{Evolving Storage Devices}
\label{subsec:evolving_devices}

The traditional storage hierarchy was historically characterized by clear and stable performance asymmetry. Mechanically driven HDDs exhibited millisecond-scale latency and limited IOPS, while DRAM provided nanosecond latency and orders-of-magnitude higher bandwidth. 
This large and consistent gap justified strict hierarchical designs in which higher tiers absorbed nearly all performance-critical requests.

However, the storage landscape has changed rapidly over the past decade. Emerging technologies—including NVMe Flash SSDs, low-latency SSDs (e.g., Optane-class devices), persistent memory (PMem), and CXL-attached memory—have significantly reduced access latency and increased internal parallelism~\cite{haas2023modern,writes-hurt,lin2023,liu2025systematic,cxl_spec_4_0}. 
Although a rough ordering in terms of access latency still exists (e.g., DRAM $<$ PMem $<$ NVMe SSD $<$ SATA SSD), bandwidth and IOPS no longer follow a strict total order across devices~\cite{haas2023modern,wu21,lin2023}.

Recent empirical studies demonstrate that performance ratios between adjacent tiers vary substantially with workload characteristics and concurrency levels~\cite{wu21,lin2023,haas2023modern}. Under low concurrency, faster devices often dominate as expected. However, under higher parallelism, internal device parallelism and queueing behavior allow lower tiers to scale, narrowing or even eliminating the throughput gap~\cite{wu21,haas2023modern}. 
For example, Optane SSD and high-end NVMe Flash SSD can exhibit nearly identical throughput under sufficiently high thread counts. In some cases, write performance ordering may even invert depending on concurrency and access patterns~\cite{wu21,writes-hurt,lin2023}.

These observations indicate the important trend. 
The modern storage hierarchy is no longer strictly hierarchical in performance; neighboring layers frequently exhibit overlapping throughput regions.
Consequently, the assumption that a higher tier strictly dominates a lower tier does not consistently hold in contemporary systems. This shift fundamentally challenges traditional hierarchical caching strategies that attempt to maximize hit rate at the fastest device. 
%-------------------------------------------------------------------------------
\subsection{NVMe and NVMe over Fabrics}
\label{subsec:nvme_of}

NVMe is a high-performance interface designed to exploit the parallelism of non-volatile memory devices over PCIe~\cite{nvme_base_spec}. Its queue-based architecture supports multiple submission and completion queues, enabling lock-free multi-core scaling and minimizing synchronization overhead~\cite{nvme_base_spec}. By allowing each CPU core to interact with independent queue pairs, NVMe reduces contention in the I/O path and sustains high IOPS and bandwidth under concurrency~\cite{nvme_base_spec,haas2023modern}.

NVMe over Fabrics (NVMe-oF) extends NVMe semantics to network transports, most notably RDMA and TCP~\cite{nvme_of_11a_hist,nvme_rdma_spec,nvme_tcp_spec}. Among these, RDMA-based NVMe-oF is particularly significant for high-performance deployments~\cite{nvme_rdma_spec}. RDMA enables zero-copy data transfer, kernel bypass, and direct memory access between hosts and remote storage targets, thereby eliminating intermediate buffer copies and reducing CPU involvement~\cite{nvme_rdma_spec}. Completion events are generated directly by the NIC, and data movement is offloaded to hardware engines, allowing NVMe command submission and completion semantics to be preserved across the network. Prior characterizations show that transport and network conditions still materially affect realized throughput and latency in disaggregated deployments~\cite{guz2017nvme,guz2018performance,kang2025understanding}.

%-------------------------------------------------------------------------------

\subsection{OpenCAS and Hybrid Storage Abstraction}
\label{subsec:opencas}

OpenCAS (Open Cache Acceleration Software)~\cite{open_cas} is an Intel-initiated open-source caching project that is now community maintained. It provides a complete block-caching stack---libraries, adapters, and tools---whose goal is to accelerate a backend block device by leveraging a higher-performance cache device. At its core is the Open CAS Framework (OCF), a high-performance caching meta-library that supports configurable caching policies. In this paper we focus on Open CAS Linux, the kernel block-layer integration used in our prototype. OpenCAS is device-agnostic: it can pair diverse cache/backend device types, including NVMe SSDs, PMem, SATA SSDs, and HDDs, while preserving a single logical volume and policy-driven routing.

As OpenCAS operates entirely within the kernel I/O path and maintains a unified fast path, it provides a practical foundation for implementing adaptive request-splitting mechanisms without modifying upper software layers.

%-------------------------------------------------------------------------------

\begin{figure}[t]
    \centering
    \includegraphics[width=0.50\textwidth]{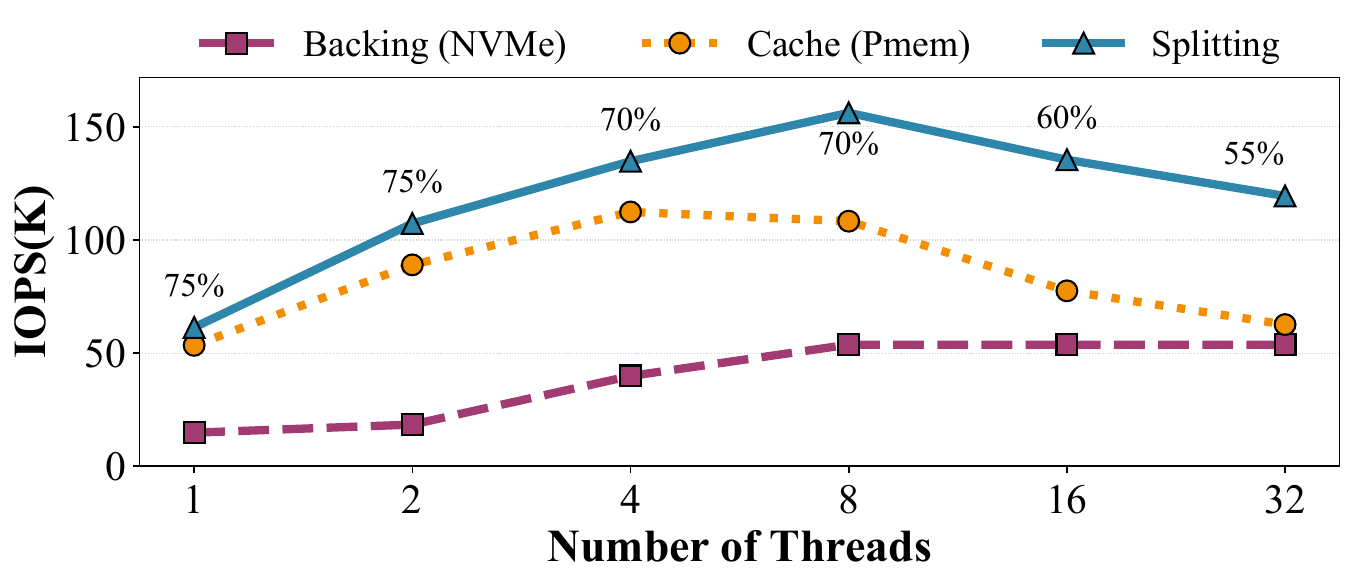}
    \caption{Throughput comparison between cache device~(PMem), backend device~(NVMe), and splitting at the optimal ratio across varying thread counts. The percentage labels on the splitting line denote the optimal split ratio at each concurrency~(e.g., 75\% indicates 75\% of requests sent to cache and 25\% to backend).}
    \label{fig:compared_to_splitting}
\end{figure}

\subsection{Non-Hierarchical Caching (NHC)}
\label{subsec:nhc}

Non-hierarchical caching (NHC), as exemplified by Orthus-style designs~\cite{wu21,lin2023}, departs from hit-rate–maximizing policies by allowing concurrent utilization of both cache and backend devices. Traditional hierarchical caching pushes for near-100\% hits: hot data are copied into the fast device, misses are promoted, and the cache tier is treated as the primary performance engine. This is appropriate when the cache is overwhelmingly faster and the backend contributes little beyond capacity. However, once the cache saturates, throughput is capped at the cache device’s peak and additional requests only add queueing; meanwhile, backend bandwidth sits idle even though it could serve useful work. This mismatch becomes more visible in modern hierarchies where device throughput curves overlap under high concurrency.

NHC treats both devices as independent contributors to total throughput and explicitly splits requests across them, even sending a fraction of cache hits to the backend when it helps. By exploiting aggregate bandwidth, performance can exceed the cache-only ceiling and approach the sum of both devices’ usable throughput when the workload and concurrency permit. In effect, NHC optimizes end-to-end throughput rather than hit ratio alone, and avoids unnecessary promotions that only add traffic to an already saturated cache.

\rev{To validate this, we perform a simple experiment: for a fixed concurrency level, we sweep the read split ratio between cache and backend (from 100\% cache to 100\% backend) and record aggregate throughput at each point.  We then compare the best-performing split against vanilla OpenCAS, which serves reads exclusively from the cache device, and against the backend device used alone.} Figure~\ref{fig:compared_to_splitting} shows the results for our cache/backend pair (PMem/NVMe). \rev{At every concurrency level, the best split exceeds both the cache-only and backend-only baselines, confirming that parallel utilization unlocks bandwidth that neither device can deliver alone.} At lower thread counts, assigning most requests to the cache is effective, but as parallelism increases the backend contributes more useful throughput and the optimal ratio shifts accordingly. This trend is consistent with prior NHC observations that throughput-optimal policies are workload- and concurrency-dependent rather than fixed~\cite{wu21}.

\subsection{Disaggregation Trends and Motivation Gap}
\label{subsec:disaggregation}
\label{sec:motivation}

Modern datacenters are increasingly shifting from server-attached disks toward \emph{disaggregated storage}, where compute and storage scale independently across the network. This shift improves hardware utilization, elasticity, and cost efficiency~\cite{zhang24dds, meta-lightning, aws-aurora}, while centralized pools simplify management and enable multi-tenancy; major cloud systems such as AWS Aurora~\cite{aws-aurora} and Azure SQL Hyperscale~\cite{azure-hyperscale} already adopt this model at scale.

NVMe over Fabrics~(NVMe-oF), introduced in \S\ref{subsec:nvme_of}, preserves key NVMe advantages even over the network~\cite{nvme_of_11a_hist}, making remote NVMe practical for latency-sensitive services. Together with high-speed fabrics~\cite{kang2025understanding, markussen24resdis}, this means remote storage can no longer be treated as an exceptionally slow tier; instead, disaggregated deployments become a realistic operating point where split policies must adapt to network variability. At the same time, \S\ref{subsec:nhc} shows that modern cache/backend pairs often have overlapping throughput regions, so maximizing hit rate is not equivalent to maximizing throughput. Figure~\ref{fig:compared_to_splitting} further shows that split routing can outperform cache-only routing and that the best split changes with concurrency. Therefore, in our setting the motivation is not to further optimize hit-rate policies, but to build a controller that continuously tracks the throughput-optimal split point.

%-------------------------------------------------------------------------------

\subsection{Challenges in Networked Environments}
\label{subsec:network_challenges}

Given the background and motivation above, the remaining gap is operational: the optimal split must be maintained under network dynamics in real deployments. In disaggregated environments, this creates the following concrete challenges:

\begin{enumerate}
\renewcommand{\labelenumi}{(\roman{enumi})}

\item \textbf{Throughput Variability.}
Backend capacity fluctuates with network congestion and cross-traffic.
A split ratio computed under one condition may quickly become suboptimal
as network state changes.

\item \textbf{Congestion Amplification.}
Over-directing requests to a temporarily degraded backend can cause queue buildup and latency inflation, which further reduces effective throughput.

\item \textbf{Imbalanced Dispatch.}
Even with a correct long-term split ratio, bursty or uneven dispatch can
cause transient device idling or saturation, reducing aggregate efficiency.

\item \textbf{Estimation Lag.}
Profiling-based or convergence-based schemes react slowly to rapid network changes, leading to persistent mismatch between assumed and actual backend performance.

\end{enumerate}

Together, these challenges motivate NetCAS as a lightweight network-aware controller that preserves NHC benefits under time-varying backend conditions.

%-------------------------------------------------------------------------------
\section{Design and Implementation}
%-------------------------------------------------------------------------------

\subsection{Framework Requirements and Design}
\begin{figure}[t]
    \centering
    \includegraphics[width=0.50\textwidth]{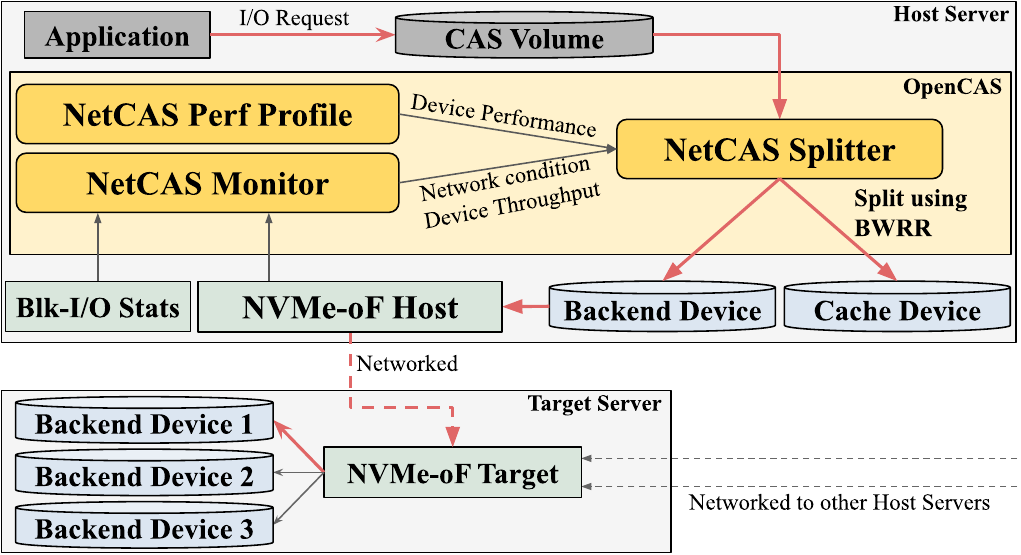}
    \caption{ NetCAS framework overview. Real time metrics from NetCAS Monitor and device baseline performance from NetCAS Perf Profile are passed to NetCAS Splitter, where I/O requests are dynamically routed to the local cache and remote backend device.}
    \label{fig:overview}
\end{figure}
Maximizing throughput in hybrid storage requires a framework that continuously adapts to device performance while remaining lightweight and transparent. NetCAS is designed to satisfy four such requirements in a unified architecture, which is depicted in Figure~\ref{fig:overview}.

\noindent
\textbf{First, real-time performance detection.} Device throughput can fluctuate significantly, especially when backend devices are remote and subject to network contention. Without timely visibility into these changes, request distribution risks either overloading a degraded device or underutilizing an available one. To enable real-time performance detection, network condition for this work, we patch the NVMe-oF host kernel module to measure fabric throughput and latency as requests complete~(\S\ref{subsec:Fluct}). By instrumenting its request-completion path, NetCAS obtains accurate, low-overhead metrics that reflect instantaneous network conditions. These metrics feed directly into the NetCAS congestion detector~(\S\ref{subsec:detect}), which quantifies bandwidth loss and latency inflation into a unified severity score for adaptation.

\noindent
\textbf{Second, adaptive splitting.} The system should adjust the split ratio in real time. Static ratios or slow convergence are inadequate when device performance or network conditions change rapidly. NetCAS relies on a precomputed Perf Profile~(\S\ref{subsec:perf_profile}) that stores empirically optimal ratios for different workload configurations, indexed by parameters. Using these parameters, NetCAS computes the ratio with the analytical model~(\S\ref{subsec:split_rat}), ensuring consistency between profiling and online adaptation.

\noindent
\textbf{Third, application transparency.} Splitting logic must operate without requiring applications or file systems to change their behavior, and must remain independent of specific caching policies~(e.g., write-back, write-through). NetCAS achieves this by extending OpenCAS. OpenCAS~\cite{open_cas} is an open-source block-level caching system that unifies a fast cache device with a slower backend device under a single logical volume, while supporting multiple caching policies. By extending OpenCAS, NetCAS can split requests across devices without altering higher-level semantics~(\S\ref{subsec:integ}).

\noindent
\textbf{Fourth, low overhead and high performance.} Additional threads, locks, or context switches could undermine throughput gains. NetCAS executes all scheduling inline with lightweight logic, avoiding extra threads or locks. Furthermore, coarse-grained request distribution, even at the correct ratio, may still stall one device while the other idles. NetCAS therefore employs Batched Weighted Round Robin~(BWRR) Scheduler~(\S\ref{subsec:bwrr}) to interleave requests, preventing blocking and keeping both cache and backend busy under high concurrency.

\subsection{Measuring Performance Fluctuations}
\label{subsec:Fluct}
In local environments, devices exhibit relatively stable performance that is easy to measure. In contrast, when backend devices are accessed over the network, fluctuations arise from congestion and resource contention, making measurement more challenging. One possible approach is a centralized controller that continuously collects global network state and allocates backend bandwidth across hosts. \rev{We intentionally avoid this design: it requires intrusive cross-node signaling, policy coupling with cluster schedulers, and tight control-plane synchronization, all of which increase deployment complexity and reaction latency.} Instead, NetCAS follows an end-host design philosophy and detects performance degradation from host-local observations.

\rev{The dominant source of variability is the network transport itself, so NetCAS instruments the NVMe-oF RDMA host kernel module to monitor throughput and latency as I/O requests complete.} Unlike coarse network counters, these metrics are \emph{storage-specific}: they reflect the actual performance of the remote device path, isolated from unrelated application traffic. \rev{These throughput and latency measurements are the \emph{sole} inputs to the congestion detector~(\S\ref{subsec:detect}), which computes the severity score used to adjust the split ratio.}

\rev{In addition to fabric metrics,} NetCAS Monitor leverages block-layer I/O counters exposed via sysfs to derive device throughput. \rev{These counters serve a complementary role: rather than feeding the congestion detector, they drive I/O detection and mode transitions~(\S\ref{subsec:integ}), determining \emph{when} to activate or deactivate splitting, but are \emph{not} used for congestion detection or split-ratio computation.}

Synchronizing these heterogeneous signals on a common sampling interval ensures that the splitter bases its decisions on both network and device conditions, while providing a modular architecture where new monitors can be integrated without restructuring the system. \rev{While our current implementation focuses on network-induced variability, the same host-local framework naturally extends to other performance-shift sources~(e.g., device-level interference or firmware throttling) by adding new metric sources.}

\rev{A natural concern is whether NetCAS must distinguish network congestion from storage-side congestion---for example, a shared storage target stuttering under garbage collection or multi-tenant load. In practice, the distinction is unnecessary: from a host's perspective, any backend slowdown appears as reduced end-to-end throughput and increased completion latency. Because NetCAS reacts to these end-to-end signals rather than diagnosing root cause, the response remains consistent regardless of origin: shift more requests to the local cache and reduce backend share. Once slowdown clears, whether network- or storage-induced, the profile-based ratio is restored immediately, avoiding the slow additive recovery of convergence-based schemes.}

\rev{Therefore, NetCAS requires no cross-node coordination protocol. Each host independently observes its own path conditions and converges to a split ratio that reflects its instantaneous fair share of backend bandwidth. This host-autonomous design keeps the data path simple, scales with cluster size, and remains deployable incrementally without introducing a global control dependency.}

\subsection{Performance Profile}
\label{subsec:perf_profile}
To split requests efficiently without incurring costly online exploration, NetCAS relies on a \textbf{Performance Profile~(Perf Profile)} that empirically records standalone throughputs of cache and backend devices for different workload configurations. The profile spans a three–dimensional space defined by \textit{block size}, \textit{in–flight requests}, and \textit{threads}. These values are later referenced when determining the optimal split ratio. Such profile–based approaches are widely used in systems for rapid decision making, as they capture device–specific scaling behaviors while enabling constant–time lookups in the fast path. By consulting a compact profile, they avoid latency and CPU overhead in storage and networking systems such as congestion–control protocols~\cite{winstein13tcp_ex_machina}, DVFS power management~\cite{liang13dvfs}, and TCP CUBIC’s cubic–root calculation~\cite{ha08cubic}.

\rev{Internally, the Perf Profile is a lookup table~(LUT) indexed by $\langle$\textit{block\_size}, \textit{inflight\_requests}, \textit{threads}$\rangle$, where each entry stores the independently measured standalone throughput of the cache device and the backend device at that operating point.  The initial grid spans 5 inflight levels $\times$ 5 thread levels $\times$ 2 block sizes $=$ 50 entries, with concurrency levels drawn from commonly exercised datacenter settings and block sizes matching the most common OpenCAS page sizes.}

\rev{These entries are populated with a \emph{generic I/O microbenchmark}~(e.g., \texttt{fio} with random reads) rather than an application-specific workload, because the LUT captures \emph{device-level concurrency characteristics}---a hardware property of each device pair---that transfer across workloads sharing the same hardware.  This is analogous to how transport-layer congestion-control profiles characterize link capacity rather than application behavior; modern datacenter environments further justify profile reuse, as hardware and workload mixes within a given cluster are highly homogeneous~\cite{mars13, nishtala13}. Industry practice confirms this pattern: Meta deploys uniform 48-rack server pods as standardized building blocks~\cite{meta_dcfabric}, Google organizes dedicated TPU Pod clusters of identical accelerators~\cite{google_tpu}, Azure assembles thousands of identical GPU servers into AI supercomputer clusters~\cite{azure_superfactory}, and Alibaba's scheduling constraints confine most jobs to a small subset of eligible nodes~\cite{cheng18alibaba}.}

\rev{When the runtime workload operates between grid points, NetCAS uses the \emph{nearest LUT entry} as a starting estimate.  The online monitor~(\S\ref{subsec:Fluct}) immediately observes actual throughput and feeds the congestion detector, which recalculates the split ratio using the formula in~\S\ref{subsec:split_rat}.  The LUT thus provides a \emph{fast starting point}; the online monitor dominates within seconds.  New entries can also be appended at runtime as the monitor encounters previously unseen workload configurations, making the profile incrementally self-improving.}

\begin{figure}[t]
  \centering
  \includegraphics[width=0.45\textwidth]{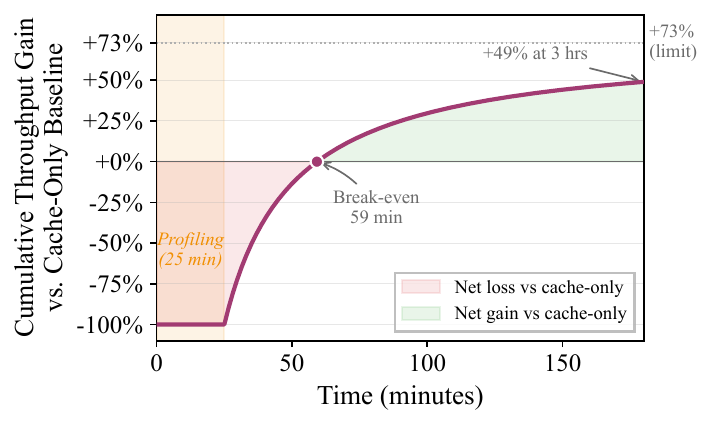}
    \caption{\rev{Cumulative throughput gain of NetCAS over a cache-only baseline (inflight requests\,=\,16, threads\,=\,16). During the one-time profiling phase~(25\,min), the system populates the LUT and operates at reduced throughput, creating an initial deficit. The break-even point is reached at 59\,min; beyond that, every additional minute of operation adds net gain, approaching the +73\% steady-state limit. At the 3-hour mark cumulative gain stands at +49\%.}}
  \label{fig:break_even_fast}

\end{figure}
Constructing the full table can be costly. In modern datacenters, however, hardware and workload mixes are relatively homogeneous, and many services operate under stable, repeatable configurations. This lets operators prebuild a profile for a fixed workload or maintain a shared profile that can be reused across servers, avoiding per–machine exploration. \rev{ Figure~\ref{fig:break_even_fast} quantifies this trade-off. The 50-entry LUT completes in approximately 25\,minutes, after which NetCAS operates at its steady-state split ratio. The cumulative gain crosses zero (break-even) at 59\,minutes and continues to grow, reaching +49\% at three hours and asymptotically approaching the +73\% steady-state advantage. Since modern datacenter jobs typically run for hours to days, the profiling cost is amortized quickly.}

\subsection{Detecting Congestion}
\label{subsec:detect}
We detect fabric anomalies using a sliding RDMA window over completed I/O to make the NetCAS detector robust to transient bursts and queuing noise. Every monitoring epoch, the NetCAS monitor exports per–epoch throughput $B_t$ and latency $L_t$. The NetCAS detector maintains baselines given by the maximum observed throughput $\bar{B}$ and the minimum observed latency $\bar{L}$, and computes normalized deviations
\[
\delta_B = \frac{\bar{B} - B_t}{\bar{B}}, \qquad
\delta_L = \frac{L_t - \bar{L}}{\bar{L}}.
\]

From these deviations we compute a single severity score
\[
\texttt{drop\_permil}
= 1000 \cdot \big(\beta_B \,\delta_B + \beta_L \,\delta_L\big),
\]
where the weights $\beta_B$ and $\beta_L$ control which signal is emphasized (set to $\beta_B=\beta_L=0.5$ in our prototype to treat bandwidth degradation and latency inflation equally). 
The result, denoted \texttt{drop\_permil}~(a per-thousand penalty factor), provides a joint view of bandwidth loss and latency inflation and is used to calculate the optimal split ratio under network congestion.
\rev{Note that our implementation can easily be extended to cache-device congestion by exposing block-layer I/O counters for the cache device without architectural changes as previously mentioned.}
%\rev{While our current implementation focuses on network-induced congestion, the same framework naturally extends to cache-device congestion by exposing block-layer I/O counters for the cache device.  Adding such a monitor requires no architectural changes; only a new metric source is plugged into the existing NetCAS Monitor.}

\subsection{Calculating the Split Ratio}
\label{subsec:split_rat}

\begin{figure}[t]
    \centering
    \includegraphics[width=0.5\textwidth]{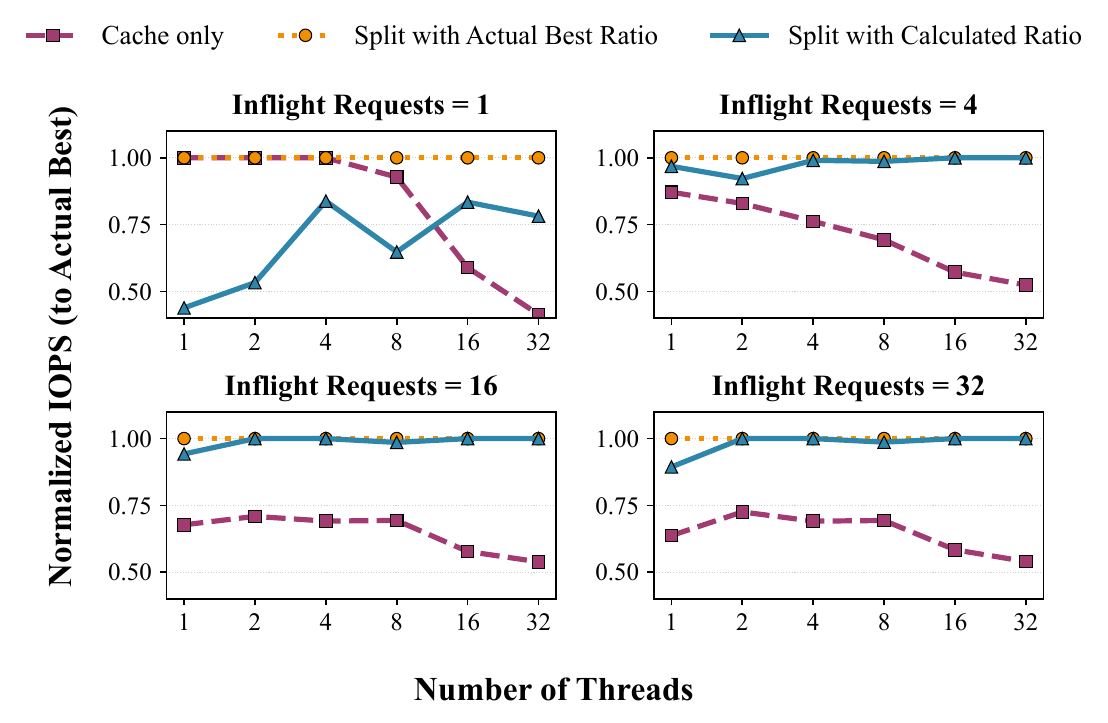}
    \caption{Normalized throughput under different inflight request counts without network congestion. At low concurrency the calculated split deviates from the empirical best, but accuracy improves quickly with higher concurrency, converging to the optimal ratio.}
    \label{fig:calculate_vs_actual}
\end{figure}

While the Perf Profile records device performances, we also need a principled way to calculate the \emph{ideal} split. Prior work~\cite{wu21} showed that when many requests are issued in parallel, completion time can be modeled by balancing the service times of each device. With a fraction $r$ of requests sent to the cache and $1-r$ to the backend, the per–device service times are
\[
T_{\text{cache}} = \tfrac{r}{I_{\text{cache}}}, \qquad
T_{\text{back}}  = \tfrac{1-r}{I_{\text{back}}},
\]
where $I_{\text{cache}}$ and $I_{\text{back}}$ are standalone throughputs from Perf Profile. The batch completes only when the slower side finishes:
\[
T_{\mathrm{total}} = \max\!\left(\tfrac{r}{I_{\text{cache}}}, \tfrac{1-r}{I_{\text{back}}}\right).
\]
The minimizer of $T_{\text{total}}$ lies at the intersection of the two, yielding $\rho_{\text{base}}$, the optimal split ratio without network congestion, 
\[
\rho_{\text{base}} = \frac{I_{\text{cache}}}{I_{\text{cache}} + I_{\text{back}}}.
\]
When congestion is detected, the splitter applies the observed \texttt{drop\_permil} $d \in [0,1000]$ to scale down the backend throughput, recomputing
\[
\rho = \frac{I_{\text{cache}}}{I_{\text{cache}} + I_{\text{back}}\!\cdot(1-d/1000)},
\]
so that the ratio adapts smoothly to degraded conditions.

In practice the model is inaccurate at very low queue depths: with only one or two in-flight requests, a single operation directed to the slower device blocks completion and variance tends to dominate. As concurrency increases, as is typical in datacenter workloads~\cite{huye2023lifting, haas2023modern, su2025dcperf}, both devices stay busy in parallel and the prediction rapidly converges to measured throughput~(Figure~\ref{fig:calculate_vs_actual}).

\subsection{Selecting the Device}
\label{subsec:bwrr}

{
\begin{algorithm}[t]
\caption{Batched Weighted Round Robin~(BWRR)}
\label{alg:bwrr}
\begin{algorithmic}[1]
\Require split\_ratio $\rho$, window\_size $W$, batch\_size $B$

\Procedure{SendCache}{$r$}
  \State cache\_quota $\gets$ cache\_quota $-1$
  \State Send $r$ to CACHE
\EndProcedure
\Procedure{SendBack}{$r$}
  \State backend\_quota $\gets$ backend\_quota $-1$
  \State Send $r$ to BACKEND
\EndProcedure

\For{each incoming req $r$}
  \If{req\_count $= W$}
    \State $a \gets \operatorname{round}(\rho W)$;\; $b \gets W-a$
    \State pattern\_size $\gets \min\!\big(W/\gcd(a,b),\,B\big)$ 
    \State pattern\_cache $\gets \big\lfloor~(pattern\_size\cdot a)/W \big\rfloor$
    \State pos $\gets 0$ ;\; req\_count $\gets 0$
    \State cache\_quota $\gets a$;\; backend\_quota $\gets b$
  \EndIf

  \If{cache\_quota $>$ 0 \textbf{and} backend\_quota $>$ 0}
    \If{pos $>$ pattern\_cache} \Call{SendBack}{r}
    \Else\; \Call{SendCache}{r}
    \EndIf
    \State pos $\gets$~(pos $+1$) $\bmod$ pattern\_size
  \ElsIf{cache\_quota $= 0$}\; \Call{SendBack}{$r$}
  \ElsIf{backend\_quota $= 0$}\;  \Call{SendCache}{$r$}
  \EndIf

  \State req\_count $\gets$ req\_count$+1$
\EndFor
\end{algorithmic}
\end{algorithm}
}

To enforce the split ratio $\rho$ in practice, NetCAS employs a \textbf{Batched Weighted Round Robin (BWRR)} scheduler. Whereas the Perf Profile stores per–device throughput for each workload—providing \emph{macroscopic} guidance by encoding the optimal long–term ratio—BWRR delivers \emph{microscopic} control at the request level so the desired ratio is realized in short windows. This prevents burstiness, avoids idle slots on either device, and keeps both cache and backend continuously utilized. Algorithm~\ref{alg:bwrr} shows the core logic.

\begin{figure}[t]
    \centering
    \includegraphics[width=0.5\textwidth]{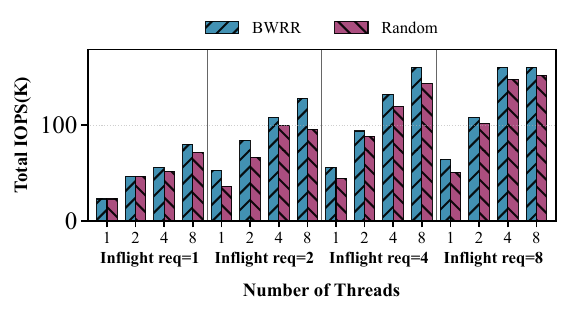}
    \caption{Throughput comparison of BWRR versus random dispatch across inflight requests and thread counts. BWRR sustains the target ratio more evenly, delivering higher aggregate IOPS especially under shallow queues where randomization causes imbalance.}
    \label{fig:BWRR_vs_Random}
\end{figure}

BWRR combines three mechanisms: (i) enforcing long–term ratios by expected counts, (ii) maintaining the ratios even within short-term intervals with minimal repeating pattern (via GCD), and (iii) filling residual imbalance with quota–based dispatch. This multi–tiered control keeps the ratio accurate even at small window sizes while avoiding bursts or starvation. As shown in the ablation study~(Fig.~\ref{fig:BWRR_vs_Random}), BWRR maintains the target ratio far more evenly than random dispatch, yielding higher aggregate throughput under shallow queues where randomization wastes parallelism.

\rev{Concretely, consider a window of $W{=}10$ requests with $\rho{=}0.7$.
BWRR assigns $a{=}7$ slots to cache and $b{=}3$ to backend.
Because $\gcd(7,3){=}1$, the minimal repeating pattern spans all 10~slots:
the first~7 go to cache, the next~3 to backend, and the cycle restarts.
As each request arrives, BWRR advances through this interleaved pattern,
guaranteeing that the target ratio is maintained even within short bursts.
If one device's quota exhausts before the window closes, all remaining
requests go to the other device, ensuring exact adherence to~$\rho$
over every window.}

\subsection{Read/Write Mixed Workload Behavior}
\label{subsec:rw_behavior}
\rev{Importantly, NetCAS does not modify, intercept, or reorder any write operation.  All splitting decisions apply exclusively to \emph{cache-hit read} requests; write allocation, eviction ordering, and durability guarantees remain exactly as implemented by the underlying OpenCAS cache mode~(write-back or write-through).  This boundary is a deliberate design choice: modifying the write path would require dependency tracking or journaling to preserve consistency~\cite{koller13}, and prior work on non-hierarchical caching likewise keeps write allocation unchanged for endurance and persistence~\cite{wu21}.  By restricting splitting to the read side, NetCAS inherits OpenCAS's proven consistency model while avoiding write-path coordination entirely.  }

\begin{figure}[t]
    \centering
    \includegraphics[width=0.45\textwidth]{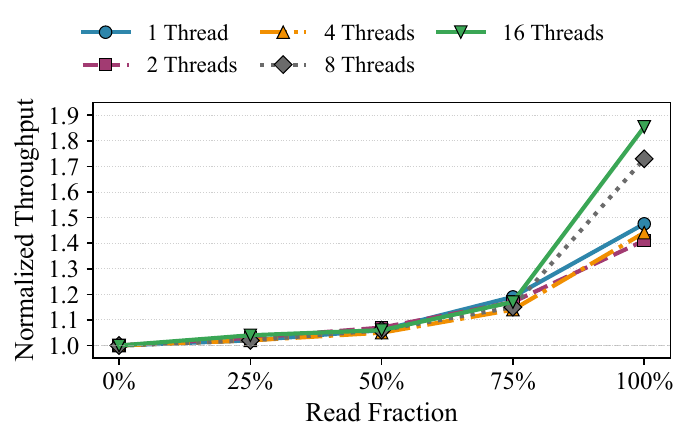}
\caption{\rev{Throughput under varying read/write ratios without network congestion (NT\,=\,16 inflights).  Gains scale roughly linearly with the read fraction but fall slightly short due to write-side contention on the cache device.}}

    \label{fig:read_write}
\end{figure}

\rev{Because NetCAS only splits cache-hit reads, its benefit scales roughly linearly with the read fraction of the workload.  Figure~\ref{fig:read_write} shows this trend across all thread counts: as the read share grows, throughput increases in near-proportion.}

\rev{The gains fall just short of perfect linear scaling at intermediate read fractions.  The gap is attributable to write traffic: in write-through mode every write must still be served exclusively by the cache device, so the cache queue never fully drains regardless of how many reads are offloaded to the backend.  This residual write contention prevents the throughput improvement from growing in exact lockstep with the read ratio.  At high thread counts (8 and 16 threads) the pure-read case reaches 1.73$\times$ and 1.85$\times$ respectively, confirming that NetCAS's benefit grows with concurrency as the splitter keeps both devices saturated.}

\subsection{Integration and Footprint}
\label{subsec:integ}

We implement splitting directly above \texttt{engine\_fast()} in OpenCAS, the unified fast path for handling cache hits across write policies. NetCAS injects NetCAS splitter here to decide whether a request is served by the cache or redirected to the backend device; misses always go to the backend device, keeping the design safe and policy-agnostic. Since allocation, completion, and locking are already handled in upper layers, the NetCAS splitter hooks into the mid-path without new threads or locks, preserving transparency while avoiding extra synchronization.

\begin{figure}[t]
    \centering
    \includegraphics[width=0.50\textwidth]{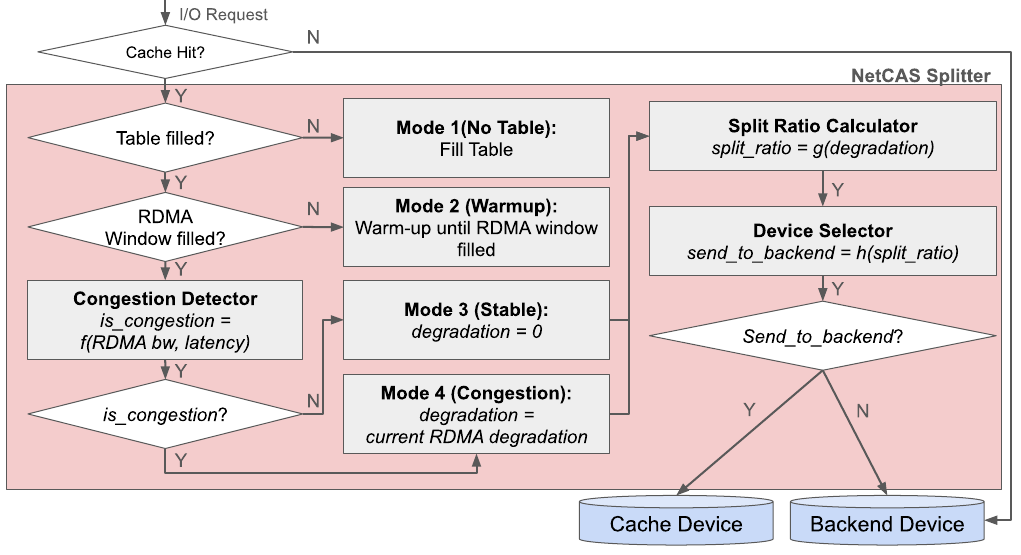}
    \caption{\rev{NetCAS mode-transition state machine. Arrows denote event-driven transitions: LUT population completes (\emph{No Table}$\to$\emph{Warmup}), monitoring baselines stabilize (\emph{Warmup}$\to$\emph{Stable}), the congestion detector fires (\emph{Stable}$\to$\emph{Congestion}), and fabric metrics recover (\emph{Congestion}$\to$\emph{Stable}). In \emph{Stable} mode the splitter uses the LUT-derived ratio with near-zero overhead; in \emph{Congestion} mode the ratio is recalculated every epoch from live network metrics.}}
    \label{fig:mode_select}
\end{figure}

To minimize overhead while adapting to network dynamics, NetCAS employs a lightweight mode-based control scheme as depicted in Figure~\ref{fig:mode_select}: \textbf{No Table} populates the Perf Profile, \textbf{Warmup} stabilizes monitoring windows, \textbf{Stable} applies precomputed ratios with near-zero cost, and \textbf{Congestion} periodically recalculates ratios to reconfigure BWRR.
\rev{Transitions are event-driven:
once the LUT is fully populated for the current workload class,
the system moves from \emph{No Table} to \emph{Warmup};
after the monitoring windows accumulate enough samples for a
statistically stable baseline, \emph{Warmup} advances to \emph{Stable}.
In \emph{Stable} mode the splitter serves requests at the LUT-derived
ratio with near-zero overhead.
When the congestion detector signals a sustained throughput drop or
latency spike on the fabric (Section~\S\ref{subsec:detect}), the
system enters \emph{Congestion} mode and recalculates the split ratio
every epoch using live network metrics.
Once fabric performance recovers, the ratio reverts to the
profile-based value and the system returns to \emph{Stable}.}
By confining calibration and monitoring to transient phases, the NetCAS splitter stays responsive to change yet incurs virtually no steady-state overhead. Even under heavy workloads (16 threads $\times$ 16 inflight requests), the total utilization of NetCAS rises only from 12.46\% (OpenCAS) to 12.79\%, a negligible 0.33\% absolute difference.

NetCAS required modest kernel modifications with less than 1,200~LOC modified to the OpenCAS and NVMe-oF RDMA kernel module. The full source code and experimental artifacts are available at: \url{https://github.com/NetCAS-SKKU/NetCAS}.
% \url{https://github.com/NetCAS-SKKU/NetCAS}.
% \url{https://github.com/netcas-anon-research/NetCAS}.
% 780 + 280

%-------------------------------------------------------------------------------
\section{Evaluation}
%-------------------------------------------------------------------------------

\subsection{Evaluation Setup}

\textbf{Testbed.}  
We evaluate NetCAS on a dual–socket Intel Xeon Gold~6330 server~(56~cores total, 2.0\,GHz) with 384\,GB DRAM running Linux~5.15 and OpenCAS~v22.3.  
\rev{OpenCAS is configured in \emph{write-through} mode so that every write is committed to the backend device synchronously; this eliminates write-ordering concerns and isolates the effect of read-path splitting.}
The cache device is a local Intel Optane Persistent Memory module, while the backend device is a remote Samsung~990~Pro NVMe SSD accessed over NVMe–oF~(RDMA) through a Mellanox ConnectX–5 100\,Gbps NIC.  
The network topology consists of three host servers connected to one target storage server through a middle switch: the target uses a 40\,Gbps NIC, while the hosts and switch use 100\,Gbps NICs, creating a single congestion point at the target.  

\textbf{Workloads and Baselines.}  
\rev{Synthetic workloads are generated with \texttt{fio} ( \texttt{bs=64k}, \texttt{ioengine=libaio}, \texttt{direct=1}, \texttt{size=1G}), sweeping I/O depth and thread count.}
\rev{For real-world application behavior, we use Filebench workloads as described in~\S\ref{subsec:real_world}, covering random-read, sequential-read, and mixed read/write patterns.}
\rev{Network congestion is injected with \texttt{ib\_write\_bw} using 1\,MB messages, queue depth~1, rate-limited to 2.5\,Gb/s.}
\rev{Before each run the cache is warmed: a sequential write fills the device, followed by 120\,s of random writes; the cache is then flushed and statistics reset.}
We compare NetCAS against three baselines: vanilla OpenCAS (cache standalone), backend standalone, and the state–of–the–art OrthusCAS~\cite{wu21}.
\rev{OrthusCAS converges its split ratio by monitoring per-device block-layer throughput statistics. Because Intel Optane PMem does not expose these counters, the convergence loop cannot operate; the ratio only re-adjusts when the cache hit rate changes. We therefore supply OrthusCAS with the empirically best static ratio for each concurrency level, giving it an upper-bound advantage that a live deployment would not achieve.}

\rev{All experiments use a prefilled, prewarmed cache so that every read hits the cache device.  This isolates the splitting mechanism: since NetCAS only modifies the read path for cache-hit requests~(\S\ref{subsec:rw_behavior}), prefilling ensures every I/O exercises the splitter.  Sections~\S\ref{subsec:baseline},~\S\ref{subsec:contention} use read-only workloads to measure the maximum splitting benefit; Section~\S\ref{subsec:rw_mixed} then introduces mixed read/write ratios to confirm that gains scale with the read fraction and that the unmodified write path introduces no regressions, consistent with the trends reported by OrthusCAS~\cite{wu21}.}

% All experiments use read–only workloads with prewarmed cache to evaluate effectiveness under cache hits.  
% Synthetic workloads are generated with \texttt{fio} using a 64\,KB block size (the page size suggested by OpenCAS). To evaluate NetCAS under contention, we inject network congestion using \texttt{ib\_write\_bw}.
% We compare NetCAS against three baselines: vanilla OpenCAS (cache standalone), backend standalone, and the state–of–the–art OrthusCAS~\cite{wu21}. For Orthus, which converges by monitoring per-device block statistics of cache and backing throughput, we instead use a statically selected split ratio since PMem lacks this interface, assuming convergence has already occurred after warmup.

\subsection{Baseline Performance}
\label{subsec:baseline}
\begin{figure}[t]
    \centering
    \includegraphics[width=0.50\textwidth]{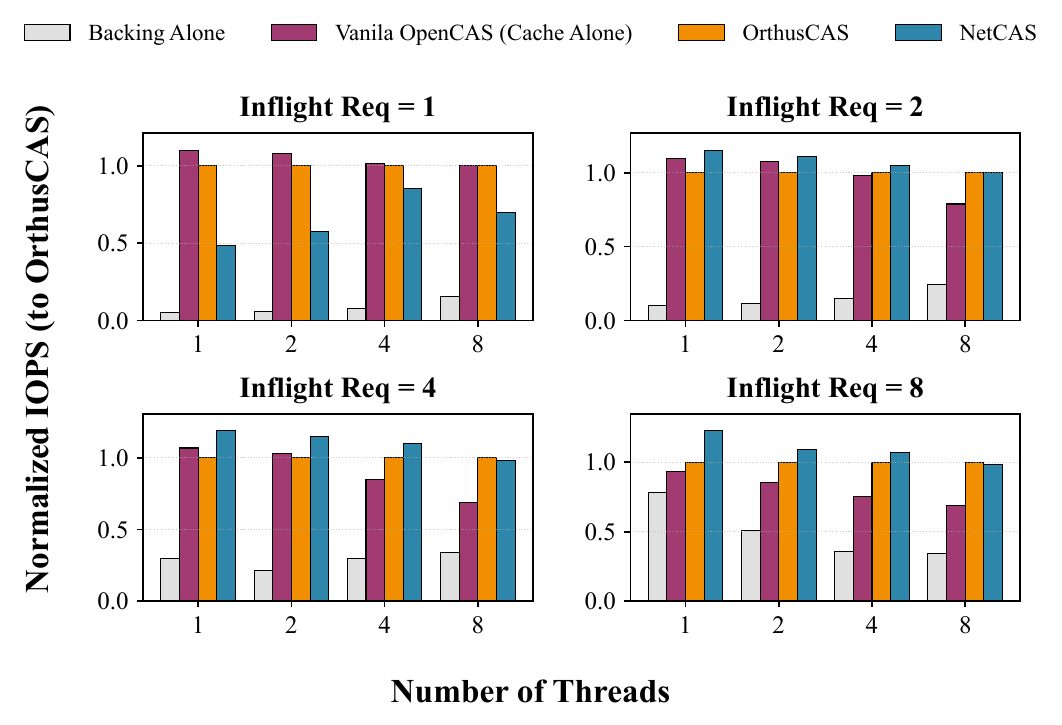}
    \caption{Baseline throughput without contention. NetCAS achieves up to 125\% higher throughput than OrthusCAS and 142\% over vanilla OpenCAS across concurrency levels.}
    \label{fig:baseline_perf}
\end{figure}
\begin{figure*}[t]
    \centering
    \includegraphics[width=0.95\textwidth]{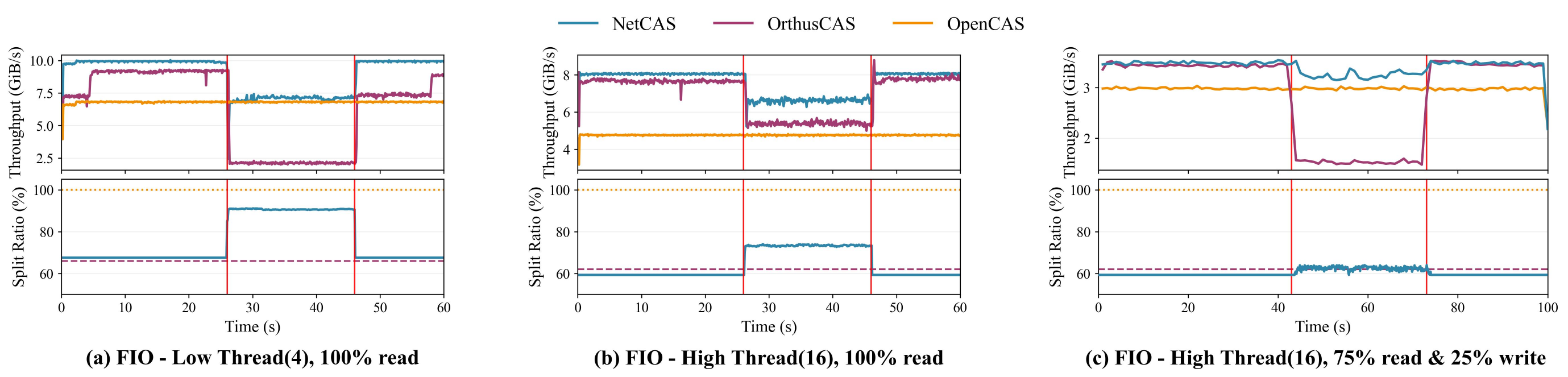}
    \caption{Throughput under injected congestion. Panels (a) and (b) use read-only \texttt{fio} with 4 and 16 threads (16 in-flight requests), with a 60\,s run and a 20\,s contention window from 10 competing flows (2 servers, 2.5\,Gbps cap per flow). Panel (c) uses mixed read/write \texttt{fio} with 16 threads and 16 in-flight requests, with a 100\,s run and a 30\,s contention window from 40 competing flows (2 servers, 2.5\,Gbps cap per flow).}
    \label{fig:throughput_congestion}
\end{figure*}

% Use normalization or Moving Average for this in LATER SUBMISSION
% \begin{figure*}[t]
%     \centering
%     \includegraphics[width=0.95\textwidth]{congestion_ab.pdf}
%     \caption{Throughput under injected congestion (20\,s). \texttt{fio} with 4 and 16 thread settings (16 inflight request each) and \texttt{TPC\-C} with 16 terminals (StockLevel 100\% for read-only workload). In all cases, contention is created by 10 competing flows from 2 servers, each capped at 2.5\,Gbps.}
%     \label{fig:throughput_congestion}
% \end{figure*}

% Low Thread - Approx 3.5X(1.7X compared to OpenCAS), High Thread - Approx 1.2X, (1.7X compared to OpenCAS)
We begin by comparing NetCAS with vanilla OpenCAS and OrthusCAS under contention-free conditions. Figure~\ref{fig:baseline_perf} reports aggregate throughput across concurrency levels. As expected, both NetCAS and OrthusCAS surpass vanilla OpenCAS by exploiting request splitting in most cases. When the number of inflight requests is one, NetCAS falls short of OrthusCAS because the analytical split ratio deviates from the empirical optimum supplied to OrthusCAS. However, by operating without extra threads or locks, NetCAS outperforms OrthusCAS in most other regimes.
\rev{At very low I/O depth~(1--2 in-flight requests), network round-trip variance dominates and a single mis-routed request can stall the pipeline, limiting the benefit of any splitting scheme. As concurrency grows, pipelining amortizes this variance and NetCAS's analytical ratio converges to the empirical optimum, explaining the widening gap in Figure~\ref{fig:baseline_perf}.}

\subsection{Performance Under Contention}
\label{subsec:contention}
% Next, we evaluate robustness under dynamic congestion. Figure~\ref{fig:throughput_congestion} shows throughput over time with congestion injected for 20s. In the synthetic \texttt{fio} benchmark (top), OrthusCAS outperforms OpenCAS under stable conditions but collapses when RDMA throughput drops. NetCAS adapts online by detecting the bandwidth loss and adjusting the split ratio, thereby sustaining high throughput. The same trend holds for TPC\-C (bottom), confirming benefits under realistic database workloads. By continuously monitoring fabric signals, NetCAS tracks instantaneous optima and avoids pathological degradation.

Next, we evaluate robustness under dynamic congestion. Figure~\ref{fig:throughput_congestion} shows throughput over time under injected load. In Figure~\ref{fig:throughput_congestion}(a) and (b), both NetCAS and OrthusCAS outperform vanilla OpenCAS by exploiting split I/O, but when RDMA bandwidth is constrained, OrthusCAS drops sharply whereas NetCAS sustains higher throughput by rebalancing online. NetCAS achieves up to 3.5$\times$ higher throughput than OrthusCAS at low-thread concurrency and still improves by about 1.2$\times$ at high-thread concurrency.

Figure~\ref{fig:throughput_congestion}(c) uses a mixed read/write \texttt{fio} workload (same thread and I/O-depth setting as (b)). Compared with (a) and (b), panel (c) with mixed read/write workload has lower aggregate throughput than read-only cases, so 10 competing flows were insufficient to induce clear contention. We therefore increased contention intensity to 40 flows each, with 100s timeline providing enough window to spawn all 40 competing flows and capture both phases consistently.

% In this case, the blue split-ratio curve is near 80\% even before external contention, although a read-only setting at the same concurrency would be around 59\%. This upward shift is expected: writes always consume backend throughput, so read-path splitting must reserve more read traffic for the cache. Because read and write requests share backend service capacity, the additional write load increases effective backend latency; NetCAS interprets this as backend pressure and naturally raises the cache share.

In the mixed read/write workload, we observe a similar trend as in the read-only cases. For vanilla OpenCAS, although write-through semantics imply that congestion should impact both reads and writes, the overall bandwidth requirement of writes in this workload is relatively low; thus, the performance degradation is dominated by reads, which are more bandwidth-intensive. As a result, both OrthusCAS and NetCAS exhibit more noticeable drops during congestion. However, NetCAS consistently maintains higher throughput. An interesting observation is that even when NetCAS dynamically adjusts its split ratio, the resulting ratio is similar to that of OrthusCAS, yet the achieved performance is significantly higher. This indicates that split ratio alone does not fully explain performance differences under mixed workloads. We attribute this gap primarily to metadata overhead. In NetCAS, read operations follow a pass-through write path that avoids metadata updates on the cache device for core-device accesses, whereas OrthusCAS updates metadata even without promoting data. Prior work has shown that metadata updates introduce non-trivial overhead~\cite{lin2023}. Because reads are more sensitive to bandwidth under congestion, this additional metadata overhead disproportionately affects OrthusCAS, leading to lower overall throughput compared to NetCAS.

% , achieving up to 2.2× higher throughput than OrthusCAS

% \rev{To verify that NetCAS preserves its advantages when writes are present, we run \texttt{fio} with mixed read/write ratios~(100/0, 70/30, 50/50, and 30/70) at 16~threads and 16~in-flight requests under the same congestion injection as~\S\ref{subsec:contention}. Figure~\ref{fig:rw_perf} reports aggregate throughput for each ratio. Because NetCAS splits only cache-hit reads while forwarding all writes through OpenCAS's unmodified write path~(\S3.1), the read-side gains diminish proportionally as the write fraction grows. At a 70/30 read/write mix, NetCAS still delivers 2.1$\times$ higher throughput than OrthusCAS under congestion; at 50/50 the advantage narrows to 1.4$\times$, and at 30/70 the three schemes converge because writes dominate and bypass the splitter entirely. These results confirm that NetCAS's read-path-only design does not introduce regressions on the write side, and that its benefits scale predictably with the read fraction of the workload—consistent with the trends reported by OrthusCAS~\cite{wu21}.toring fabric signals, NetCAS tracks the instantaneous balance point and avoids pathological degradation.}

\begin{figure}[t]
    \centering
    \includegraphics[width=0.50\textwidth]{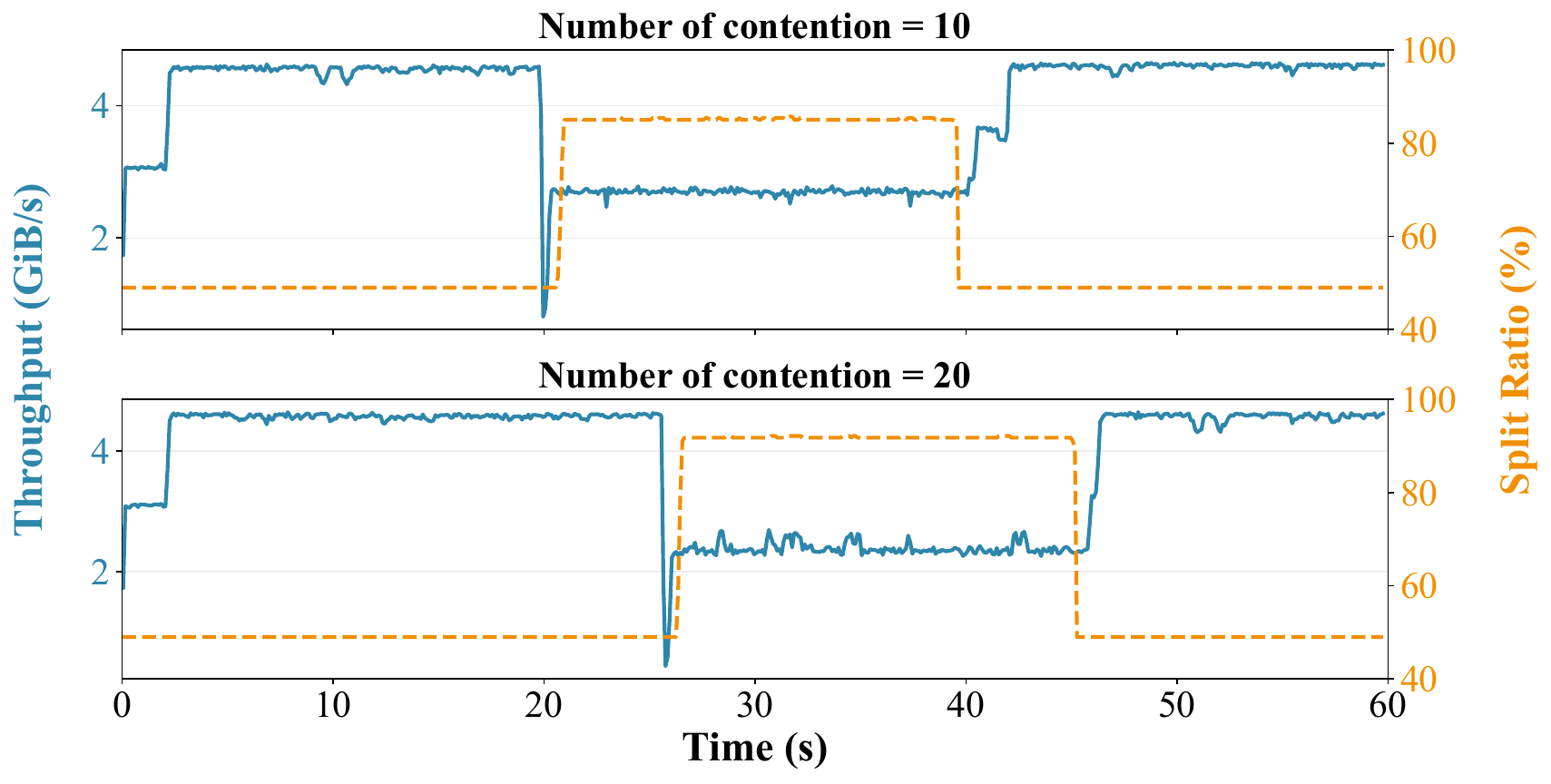}
    \caption{Throughput under low and high contention for the same workload (inflight requests = 16, threads = 16). Each competing flow attempts to maximize its bandwidth without capping. NetCAS allocates a larger share to the cache as backend bandwidth is constrained, mitigating throughput loss.}
    \label{fig:multi_congestion}
\end{figure}

% \subsection{Performance With Read \& Write Mixed Workloads}

\subsection{Performance Across Contention Levels}
Since networked storage must serve workloads under diverse levels of background competition, we conduct an experiment to evaluate NetCAS across different contention loads (i.e., varying numbers of competing flows). Figure~\ref{fig:multi_congestion} shows the resulting throughput. As backend bandwidth is squeezed by more contenders, NetCAS splitter raises the cache share, defending overall throughput without abrupt shifts. This smooth rebalancing demonstrates NetCAS’s ability to scale protection gracefully under both light and heavy competition, rather than oscillating between extremes. 
%By continuously monitoring backend signals, NetCAS tracks the shifting balance point and sustains near-optimal performance across contention levels.  
This adaptability is especially relevant in datacenter networks where tenant bandwidth demands vary dynamically. By adjusting smoothly across contention levels, NetCAS avoids cliff effects and delivers high and predictable performance. 
At the same time, it preserves fairness: instead of monopolizing backend bandwidth, it stabilizes throughput by shifting excess load to the cache while respecting each flow’s fabric share.

\subsection{Performance With Real-World Workloads}
\label{subsec:real_world}

To evaluate NetCAS under realistic application behavior, we use \textit{Filebench}~\cite{tarasov16filebench,filebench_wiki}, an application-level storage benchmark. Filebench generates workloads at the file-system interface, capturing key characteristics of real applications such as access locality, concurrency, and mixed I/O behavior.

We construct three representative workloads over a 10\,GB dataset (1000 files, each 10\,MB). Before each workload run, we apply the same warmup procedure to preload cache state for a consistent starting condition. All workloads bypass the page cache using \texttt{directio} for reads to directly exercise the cache-backend split path.

\textbf{Workload A} consists of 16 concurrent reader threads issuing 64\,KB random reads over the dataset, representing a cache-friendly scenario that isolates cache-hit performance.

\textbf{Workload B} uses 16 threads performing sequential reads with 1\,MB I/O size, scanning entire files. This models streaming or scan-heavy applications (e.g., analytics), where temporal locality is minimal and caching benefits are limited.

\textbf{Workload C} is a mixed workload with 16 reader threads and 2 writer threads. Readers perform 64\,KB random reads with \texttt{directio}, while writers issue buffered random writes to maintain system stability. This setup emulates a read-dominant service workload with concurrent read/write interference.

\label{subsec:rw_mixed}
\begin{figure}[t]
    \centering
    \includegraphics[width=0.50\textwidth]{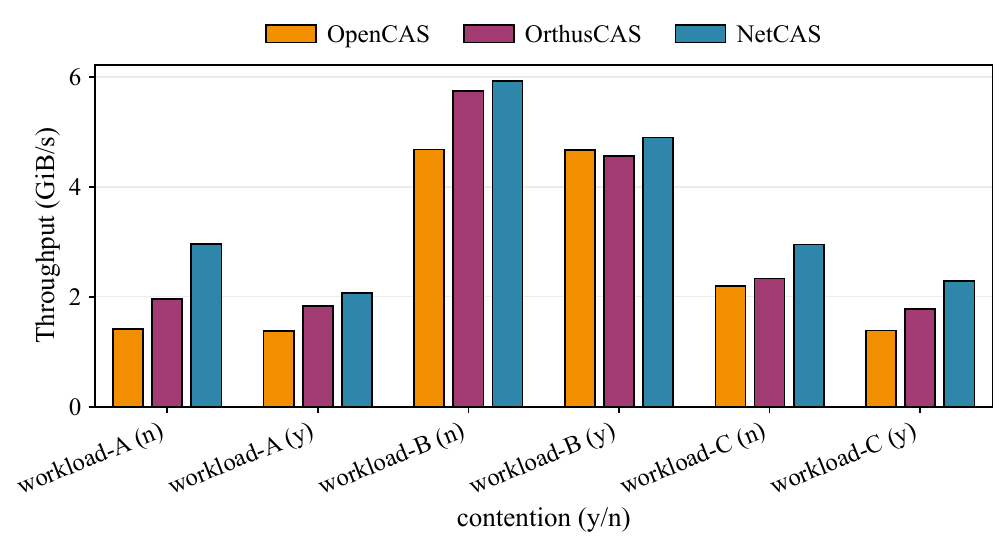}
    \caption{Throughput of Filebench workloads A–C. Labels (n) and (y) indicate without and with contention, respectively; contested runs use 40 flows (two servers, 2.5Gbps cap per flow).}
    \label{fig:filebench_bar}
\end{figure}

% Figure~\ref{fig:filebench_bar} shows throughput across these workloads, with and without contention. Under stable conditions, NetCAS consistently achieves the highest throughput, with the largest gain in Workload~A due to efficient cache-hit handling. For Workload~B, the gap is smaller as caching benefits are inherently limited, yet NetCAS incurs no additional overhead. \rev{For Workload~C, which mixes reads and writes, even Vanilla OpenCAS sees a
% throughput drop under contention: writes are pass-through operations that must traverse the congested network path regardless of caching policy.} However, NetCAS sustains the highest throughput across all workloads. In particular, OrthusCAS is more affected by degraded backend performance, while NetCAS adapts its split to mitigate the impact.

Figure~\ref{fig:filebench_bar} shows throughput across these workloads, with and without contention.
Under stable conditions, NetCAS consistently achieves the highest throughput, 
with the largest gain in Workload~A\rev{, 2.1$\times$ over OpenCAS and
1.5$\times$ over OrthusCAS,} due to efficient cache-hit handling.
\rev{For Workload~C, which mixes reads and writes, even vanilla OpenCAS sees a
37\% throughput drop under contention: writes are pass-through operations that
must traverse the congested network path regardless of caching policy.}
However, NetCAS sustains the highest throughput across all workloads\rev{, 1.65$\times$
over OpenCAS and 1.29$\times$ over OrthusCAS on Workload~C under contention}.
In particular, OrthusCAS is more affected by degraded backend performance,
while NetCAS adapts its split to mitigate the impact.

\begin{figure}[t]
    \centering
    \includegraphics[width=0.50\textwidth]{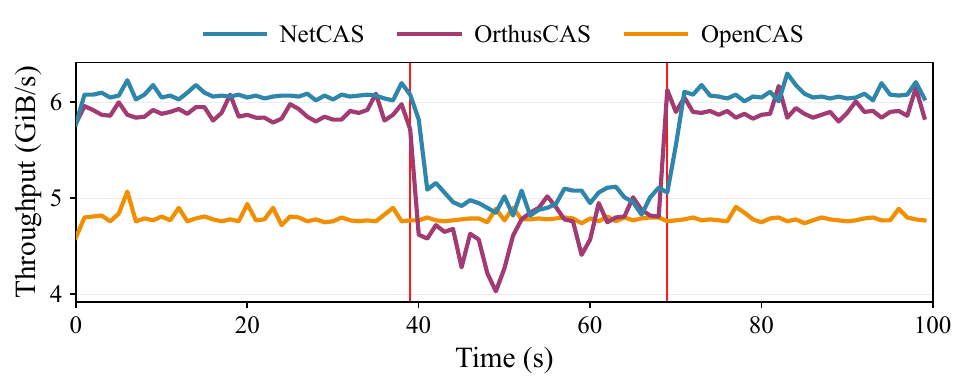}
    \caption{Throughput under injected congestion (30s). Filebench seqread (Workload B) throughput over time for the three systems; the highlighted interval corresponds to 30 seconds of RDMA contention from 40 flows (two servers, 2.5Gbps cap per flow).}
    \label{fig:filebench_workloadB}
\end{figure}

Figure~\ref{fig:filebench_workloadB} shows time-series throughput for Workload~B under a 30\,s congestion interval. OpenCAS serves all cache-hit reads from the local device and is therefore unaffected by network congestion, but it remains at a lower throughput ceiling because it cannot leverage the backend's additional bandwidth (\rev{NetCAS achieves 1.27$\times$ OpenCAS throughput in steady state}). Once congestion begins, OrthusCAS exhibits pronounced throughput fluctuations: its static split ratio continues directing a fixed fraction of reads to the now-congested backend, \rev{causing a 20\% throughput drop that pushes it below even vanilla OpenCAS}. NetCAS adaptively shifts its ratio toward the cache, \rev{limiting its throughput reduction to 17\% and sustaining 1.07$\times$ the throughput of OrthusCAS during the congestion window,} and quickly recovers after congestion clears, restoring steady-state performance within a short period. These results confirm that NetCAS's adaptive splitting remains effective under realistic file-system workloads and dynamic network conditions.

%-------------------------------------------------------------------------------
\section{Conclusion}
%-------------------------------------------------------------------------------

This paper presented NetCAS, a lightweight yet practical framework for hybrid storage in disaggregated environments, where cache and backend devices should be used \emph{concurrently} rather than strictly hierarchically. The key challenge in this setting is that backend performance is no longer stable: even with NVMe-oF and RDMA, shared-fabric contention can rapidly shift the throughput-optimal split point. NetCAS addresses this by combining three components in a unified fast path: (i) a precomputed Perf Profile that captures device-level scaling characteristics across workload configurations, (ii) host-local end-to-end monitoring of throughput and latency from the NVMe-oF completion path, and (iii) low-overhead split enforcement with batched weighted round robin. Together, these components allow each host to adapt quickly to time-varying conditions without depending on a centralized control plane or cross-node coordination protocol.

% %-------------------------------------------------------------------------------
% \section{Acknowledgement}
% %-------------------------------------------------------------------------------

% \rev{TODO TODO TODO TODO TODO TODO TODO TODO TODO TODO TODO TODO TODO TODO TODO TODO TODO TODO TODO TODO TODO TODO TODO TODO TODO TODO TODO TODO TODO TODO TODO TODO TODO TODO TODO TODO TODO TODO TODO TODO TODO TODO TODO TODO TODO TODO TODO TODO TODO TODO TODO TODO TODO TODO TODO TODO TODO TODO TODO TODO TODO TODO TODO TODO TODO TODO TODO TODO TODO TODO TODO TODO TODO TODO TODO TODO TODO  }
% %-------------------------------------------------------------------------------
\newpage

\bibliographystyle{IEEEtran}
\bibliography{references}

\end{document}